\documentclass[10pt]{llncs}
\usepackage{verbatim}
\usepackage[TS1,T1]{fontenc}
\newcommand{\tsone}{\fontencoding{TS1}\selectfont}
\usepackage[dvipdf]{graphicx}
\usepackage{latexsym}
\usepackage{xspace}
\usepackage{subfigure}

\usepackage{amsmath}
\usepackage{amssymb}
\newcommand{\codename}{{Emmerald}}

\title{General Matrix-Matrix Multiplication using SIMD features of the PIII}

\author{Douglas Aberdeen \and Jonathan Baxter\\
}

\institute{Research School of Information Sciences and Engineering \\
Australian National University\\
}

\begin{document}
\maketitle
\begin{abstract}

Generalised matrix-matrix multiplication forms the kernel of many 
mathematical algorithms. A faster matrix-matrix multiply immediately 
benefits these algorithms.  In this paper we implement efficient 
matrix multiplication for large matrices using the 
floating point Intel Pentium SIMD (Single Instruction Multiple Data) 
architecture.
A description of the issues and our solution is presented, paying
attention to all levels of the memory hierarchy. Our results demonstrate an
average performance of 2.09 times faster than the leading public domain 
matrix-matrix multiply routines.

\end{abstract}
 
\section{Introduction}

A range of applications such as artificial neural networks benefit from GEMM
(generalised matrix-matrix) multiply routines that run as fast as possible. The
challenge is to use the CPUs peak floating point performance when memory access
is fundamentally slow. The SSE (SIMD Streaming Extensions) 
instructions of the Intel Pentium III chips allow four 32-bit 
(single precision) floating point operations to be performed 
simultaneously. Consequently, efficient use of the memory 
hierarchy is critical to being able to supply
data fast enough to keep the CPU fully utilised.  In this paper we 
focus on the implementation of an efficient algorithm for the Pentium 
SIMD architecture to achieve fast, large matrix-matrix multiplies. 
Our code has been nicknamed {\codename}. 

Without resorting to the complexities associated with implementing
Strassen's algorithm on deep-memory hierarchy machines
\cite{thottethodi:1998}, dense matrix-matrix
multiplication requires $2MNK$ floating point operations where $A:M
\times K$ and $B:K \times N$ define the dimensions of the two
matrices. Although this complexity is fixed, skillful use of the
memory hierarchy can dramatically reduce overheads not directly
associated with floating point operations. It is the optimization of
memory hierarchy combined with the SSE that gives Emmerald its performance.

Emmerald implements the SGEMM interface of Level-3 BLAS, and so may be
used immediately to improve the performance of single-precision
libraries based on BLAS (such as LAPACK \cite{blas:1998}).  There have
been several recent attempts at automatic optimization of GEMM for
deep-memory hierarchy machines, most notable are PHiPAC
\cite{bilmes:1996} and the more recent ATLAS \cite{whaley:2000}.
ATLAS in particular achieves performance close to vendor optimized
commercial GEMMs. Neither ATLAS nor PhiPAC make use if the SSE
instructions on the PIII for their implementation of SGEMM.

Our experiments showed 
that ATLAS achieves a peak of 375 MFlops/s for single-precision 
multiplies on a PIII @ 450 MHz, or $0.83 \times${\em clock rate}. 
Our matrix-matrix multiply using 
SIMD instructions achieves a peak of 890 MFlops/s, or 
$1.98 \times${\em clock rate}. We also report an 
application with a price/performance ratio under USD 
\$1/MFlop/s. The following section will describe our novel use of the SSE
for {\codename}, followed by a description of optimizations designed to
improve use of the memory hierarchy. The papers concludes with a comparison
of results between ATLAS and {\codename}.

\section{SIMD Parallelisation}
\label{SIMD Parallelisation}

\begin{figure}
\centering
\mbox{
\subfigure[{Allocation of the eight SSE registers (\texttt{xmm[0-7]}), 
showing progression of the dot
products which form the innermost loop of the algorithm. Each 
circle is an element in the matrix.  Each dashed square
represents one floating point value in an SSE register.}]{
\includegraphics[scale=0.55]{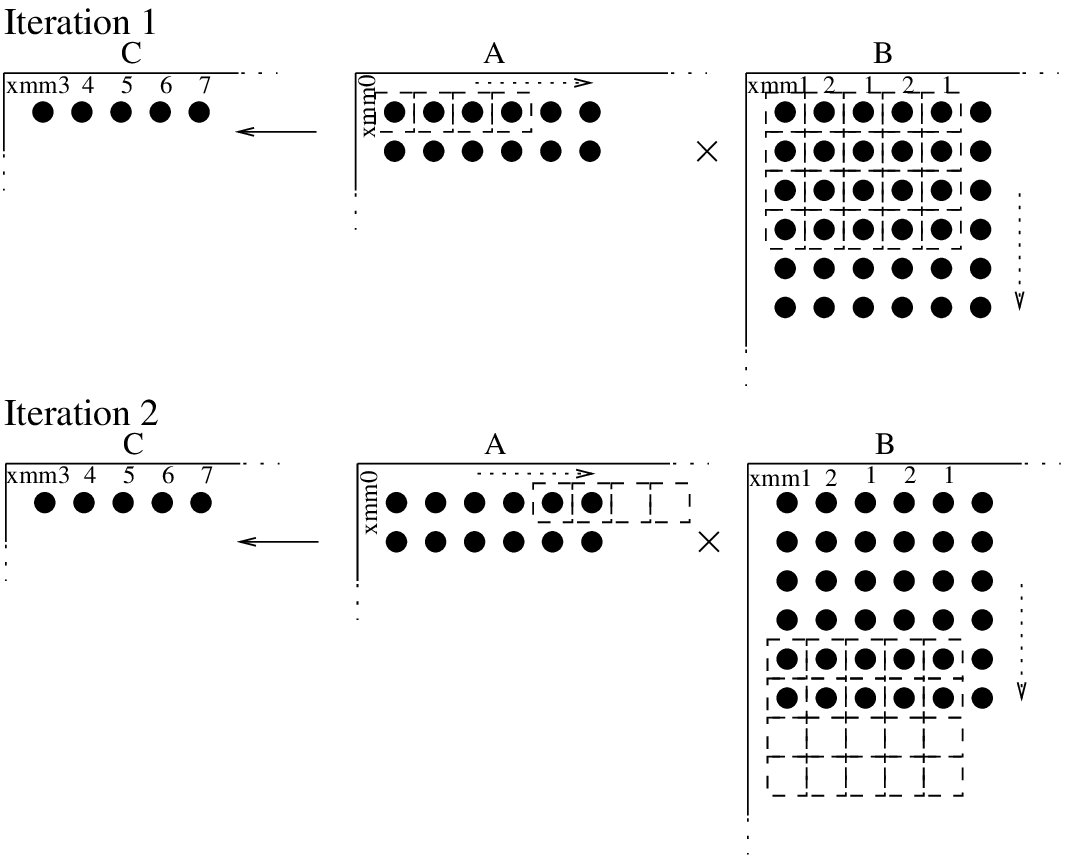}
\label{f:iter}
}
\subfigure[{L1 blocking for \codename: $C' \leftarrow A'B'$ where $A'$ and $B'$
are in L1 and $C'$ is accumulated in registers.}]{
\includegraphics[height=4cm,width=5cm]{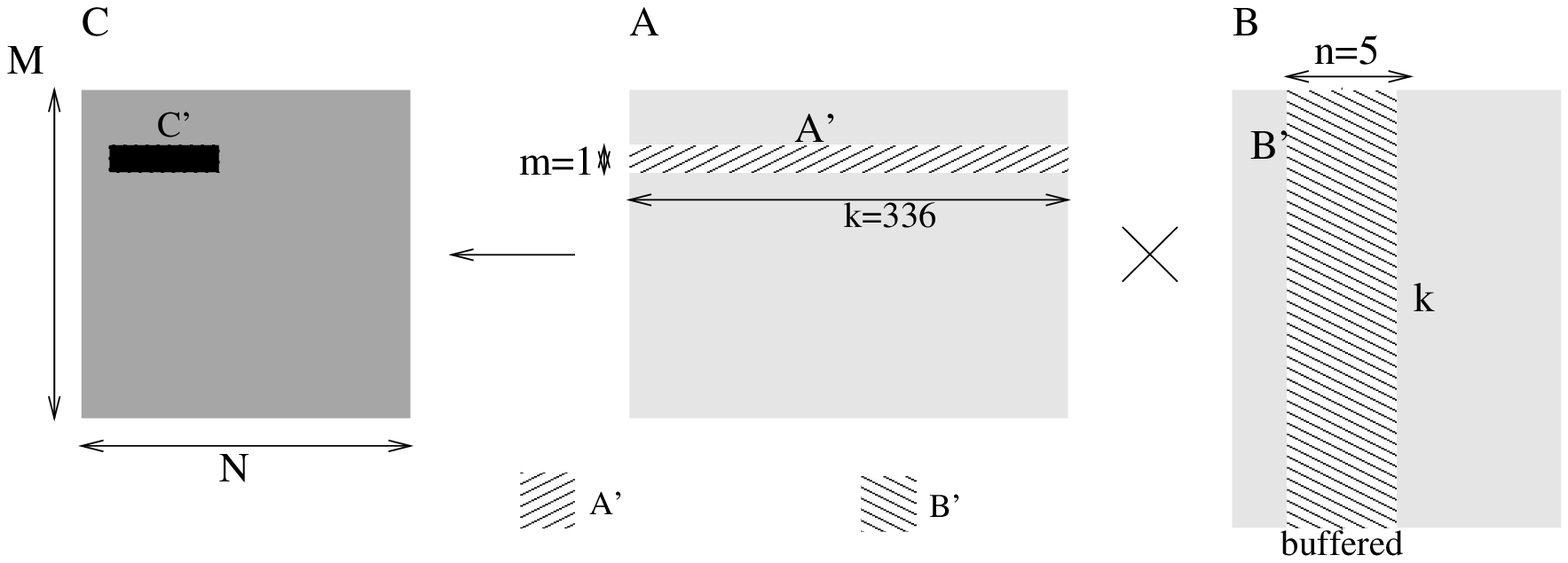}
\label{f:l1cache}
}}
\end{figure}

Two core strategies are employed to minimise 
the ratio of memory accesses to floating point operations: 
accumulate results in registers for as long as possible 
to reduce write backs, and re-use values in registers as much as possible.
In \cite{greer:1997} several dot-products were performed in parallel
as the innermost loop of the GEMM.  We took the same approach and
found experimentally that 5 dot-products in the inner loop gave the
best performance.  Figure \ref{f:iter} shows how these 5 dot products
utilise SIMD parallelism.

Four values from a row of $A$ are
loaded into an SSE register. This is re-used five times by
doing SIMD multiplication with four values from each of the five columns 
of $B$ used in the inner loop. Two SSE registers are allocated to 
loading values from $B$. Results are accumulated into the 
remaining five SSE registers. When the
dot-product ends each SSE result 
register contains four partial dot-product sums. These are summed with 
each other then written back to memory. For the best efficiency, the 
dot product length is maximised with the constraint that all data 
must fit into  L1 cache. 

\stepcounter{figure}
\section{Memory Hierarchy Optimizations}

A number of standard techniques are used in \codename\ to improve
performance. Briefly, they include:
\begin{itemize}
\item \emph{L1 blocking}:
\codename uses matrix blocking \cite{greer:1997,bilmes:1996,whaley:2000} to
ensure the inner loop is operating on data in L1 cache.
Figure~\ref{f:l1cache} shows the L1 blocking scheme. The block dimensions
$m$ and $n$ are determined by the configuration of dot-products in the 
inner loop 
(Section~\ref{SIMD Parallelisation}) and $k$ was determined experimentally.
\item \emph{Unrolling}: The innermost loop is completely unrolled 
for all possible lengths of 
$k$ in L1 cache blocks, taking care to avoid overflowing the 
instruction cache. 
\item \emph{Re-buffering}:
Since $B'$ is large $(336 \times 5)$ compared to
$A'$ 
$(1 \times 336)$, we deliberately buffer $B'$ into L1 cache. 
By also re-ordering $B$ to enforce optimal memory access patterns
we minimise translation look-aside buffer misses \cite{whaley:2000}.
\item \emph{Pre-fetching}: Values from $A'$ are not 
in L1 cache. We make use of SSE pre-fetch assembler instructions  
to bring $A'$ values into L1 cache when needed.
\item \emph{L2 Blocking}: Efficient L2
cache blocking ensures that peak rates can be maintained as 
long as $A$, $B$ and $C$ fit into main memory.
\end{itemize}

\section{Results}
\label{Results}
The performance of \codename\ was measured by timing matrix multiply
calls with size $M=N=K=16$ up to 700. 
The following steps were taken to ensure a conservative performance
estimate: wall clock time on an unloaded machine is used rather the CPU 
time; the stride of the matrices (which determines the separation in 
memory between each row of matrix data) is fixed to 700 rather than the 
length of the row; caches are flushed between calls to \texttt{sgemm()}.
Timings were performed on a PIII 450 MHz running Linux (kernel 2.2.12).

\begin{figure}
\begin{center}
\includegraphics[width=10cm,height=5cm]{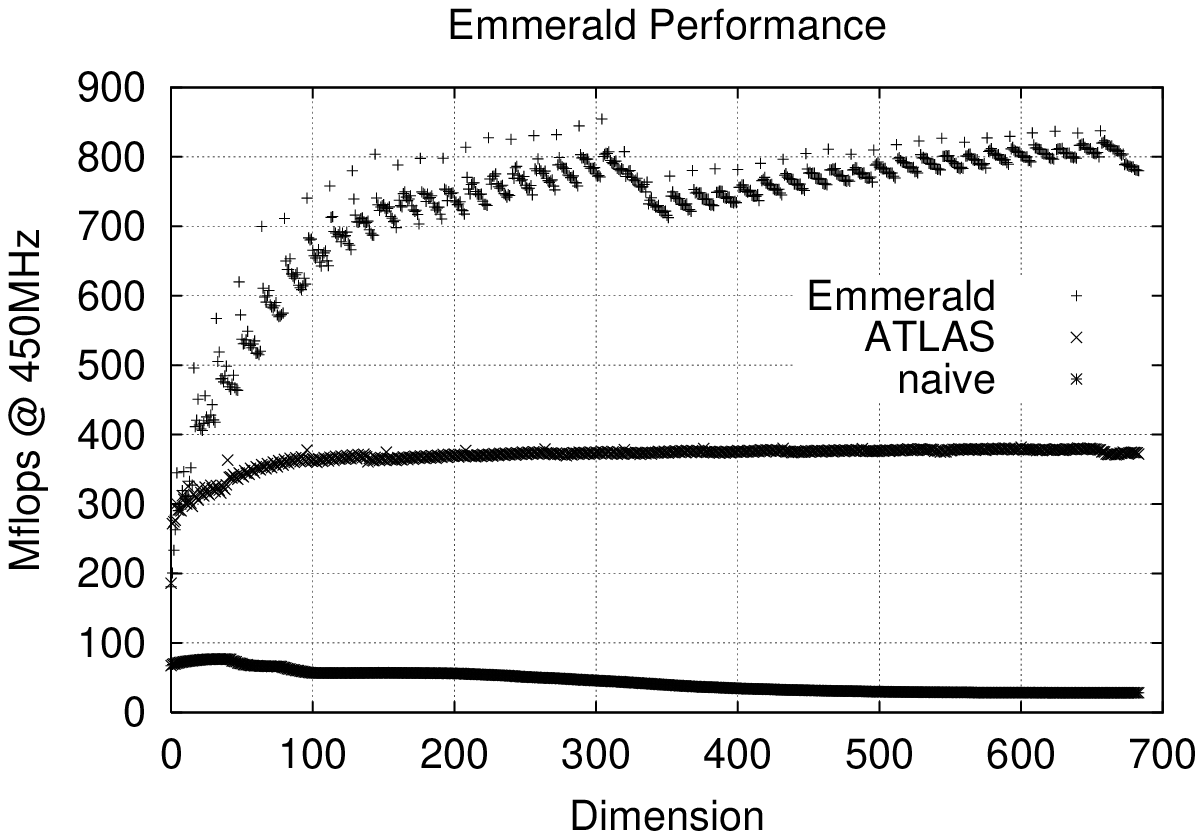}
\caption{Performance of Emmerald on a PIII running at 450 MHz compared
to ATLAS sgemm and a naive 3-loop matrix multiply. Note that ATLAS does not
make use of the PIII SSE instructions.}
\label{f:results}
\end{center}
\end{figure}

Figure~\ref{f:results} shows {\codename}'s performance compared to
ATLAS and a naive three-loop matrix multiply. 
The average MFlop/s rate of \codename\ after size
100 is 1.69 times the clock rate of the processor 
and 2.09 times faster than ATLAS. 
A peak rate of 890 MFlops/s is achieved when $m=n=k=stride=320$. 
This represents 1.97 times the clock rate.
The largest tested size was $m=n=k=stride=3696$ on a 550 MHz machine 
which ran at 940 MFlops/s. 

We have used \codename\ in distributed 
training of large Neural Networks with more
than one million adjustable parameters and a similar number of
training examples \cite{aberdeen-II:2000}. By distributing training over 196 
Intel Pentium III 550 MHz processors, and using \codename\ as the 
kernel of the training
procedure, we achieved a sustained performance of 
152GFlops/s for a price performance ratio of 98{\tsone ¢} 
USD/MFlop/s (single precision).

\section{Conclusion}
\label{Conclusion}

This paper has presented {\codename}, a version of SGEMM that utilises the 
SIMD instructions and cache hierarchy of the Intel PIII architecture. An 
application demonstrating the cost-effectiveness of such work was also 
reported. 
This code and the full version of this paper is available 
from  
\texttt{http://beaker.anu.edu.au/ research.html}.

\small
\bibliographystyle{abbrv}
\bibliography{papers}

\end{document}